# Beyond Fresnel Wave Surfaces: Off-Shell Photonic Density of States and Near-Fields in Isotropy-Broken Materials with Loss or Gain.

Maxim Durach*, David Keene

Center for Advanced Materials Science, Department of Biochemistry, Chemistry and Physics, Georgia Southern University, Statesboro, GA 30460, USA

* correspondence: mdurach@georgiasouthern.edu

Abstract: Fresnel wave surfaces, or isofrequency light shells, provide a powerful framework for describing electromagnetic wave propagation in anisotropic media, yet their applicability is restricted to reciprocal, lossless materials and far-field radiation. This paper extends the concept by incorporating near-field effects and non-Hermitian responses arising in media with loss, gain, or non-reciprocity. Using the Om-potential approach to macroscopic electromagnetism, we reinterpret near fields as *off-shell* electromagnetic modes, in analogy with off-shell states in quantum field theory. We show that photonic density of states (PDOS) distributions near Fresnel surfaces acquire Lorentzian broadening in non-reciprocal media, directly linking this effect to the Beer–Bouguer–Lambert law of exponential attenuation or amplification. Furthermore, we demonstrate how Abraham and Minkowski momenta, locked to light shells in the far field, naturally shift to characterize source structures in the near-field regime. This unified treatment bridges the gap between sources and radiation, on-shell and off-shell modes, and reciprocal and non-reciprocal responses. The framework provides both fundamental insight into structured light and practical tools for the design of emitters and metamaterial platforms relevant to emerging technologies such as 6G communications, photonic density-of-states engineering, and non-Hermitian photonics.

1. Introduction

The Fourth Industrial Revolution (4IR) seeks to integrate the physical, chemical, biological, and digital domains through their common electromagnetic foundation [1]. At the most fundamental level, systems ranging from atoms and living cells to magnetoplasmas can be understood as electromagnetic charge–field systems. Recent work introducing the "Om" ॐ -theory of macroscopic electromagnetism has proposed the Om-potential as a framework that unifies sources and fields, providing a universal language that can connect natural processes, technological systems, and even philosophical models [2].

In practice, 4IR is currently driven by the demand for advanced communications technologies enabling artificial intelligence (AI), machine-to-machine interaction, the Internet of Things (IoT), swarm robotics, and related capabilities [3]. Overcoming existing limitations will require conceptual advances comparable in impact to the steam engine, alternating-current power distribution, or semiconductors. The Om-potential approach suggests one such advance: rethinking the holistic design of emitters and structured electromagnetic fields embedded in anisotropic or non-Hermitian materials. This perspective aligns with strategies proposed for next-generation (e.g., 6G) telecommunications, which emphasize phased arrays, mode multiplexing, structured light for attenuation compensation, and the use of metamaterials along with non-reciprocal or non-Hermitian media [4-10]

Classical electromagnetic theory distinguishes near fields and far fields via their reactive and radiative components. In lossless isotropic media, far fields decay as $1/r$, reflecting intensity divergence over solid angle [11, 12]. These far fields are solutions of the homogeneous Maxwell equations, representing radiation in the absence of sources. Conceptually, they correspond to placing sources at infinity and assuming propagation without scattering or absorption. By contrast, in media with loss or gain, far fields are modulated by exponential attenuation or amplification described by the Beer–Bouguer–Lambert (BBL) law [13-15].

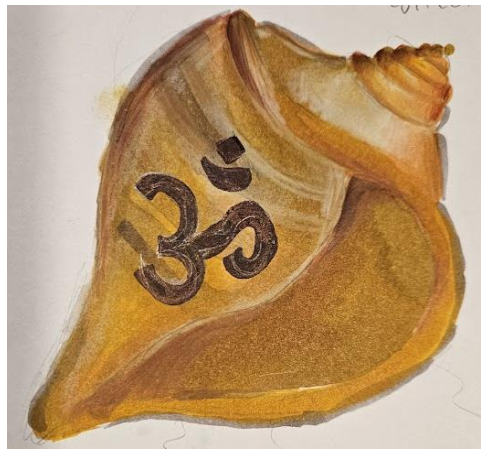

Fig. 1. Metaphorical illustration. The conch's surface represents the isofrequency light shell, while its finite wall thickness symbolizes the Lorentzian broadening from loss. The ॐ mark denotes the Om-potential, linking the off-shell near-fields to the radiating on-shell surface.

Isofrequency surfaces - also known as Fresnel wave surfaces or light shells - encode the topological features of dispersion in momentum ($k$) space, playing a central role in anisotropic and hyperbolic media [16, 17]. They are closely connected to phenomena such as modified photonic density of states (PDOS), Purcell-type emission enhancements [18, 19], and the distinction between Abraham and Minkowski momenta, which arise from refraction and ray-wave tilt [20-22]. However, Fresnel wave surfaces have been fully developed only for reciprocal, lossless materials and strictly capture far-field behavior. Near fields, in contrast, correspond to solutions of inhomogeneous Maxwell equations, and thus lie outside this framework - leaving a key theoretical gap directly addressed by ॐ-theory (see Fig. 1 for metaphoric illustration) [2].

In non-reciprocal or non-Hermitian media, fields undergo exponential attenuation or amplification in accordance with the BBL law, a phenomenon of particular interest for structured light carrying orbital angular momentum (OAM) [23, 24]. Yet this behavior cannot be represented within the conventional Fresnel-wave-surface formalism. A new framework is therefore required—one that retains the descriptive power of light shells while extending naturally to near-field regimes and non-Hermitian responses.

In this paper, we bridge that gap by directly linking Fresnel wave surfaces with PDOS in momentum space, showing that direction-dependent BBL extinction or amplification can be modeled as Lorentzian broadening of the surfaces. Within the Om-potential framework, we interpret *off-shell* electromagnetic modes as near fields, and we demonstrate that quantities traditionally associated with light shells in the far field - such as Abraham and Minkowski momenta and wavefront curvature - shift naturally toward characterizing source structure in the near-field regime.

## 2. ॐ-theory

In the presence of electric and magnetic sources $\boldsymbol{j}_e$ and $\boldsymbol{j}_m$, macroscopic electromagnetic fields satisfy Maxwell's equations

$$\nabla \times \boldsymbol{H} - \frac{1}{c}\frac{\partial \boldsymbol{D}}{\partial t} = \frac{4\pi}{c}\boldsymbol{j}_e, \quad -\nabla \times \boldsymbol{E} - \frac{1}{c}\frac{\partial \boldsymbol{B}}{\partial t} = \frac{4\pi}{c}\boldsymbol{j}_m \tag{1}$$

The constitutive relations are generally expressed as

$$\begin{pmatrix}\boldsymbol{D}\\ \boldsymbol{B}\end{pmatrix} = \widehat{M}\begin{pmatrix}\boldsymbol{E}\\ \boldsymbol{H}\end{pmatrix} = \begin{pmatrix}\hat{\epsilon} & \widehat{X}\\ \widehat{Y} & \hat{\mu}\end{pmatrix}\begin{pmatrix}\boldsymbol{E}\\ \boldsymbol{H}\end{pmatrix} \tag{2}$$

Equations (1) and (2) can be combined into the matrix form

$$\left(\widehat{Q} - \frac{1}{c}\frac{\partial}{\partial t}\widehat{M}\right)\begin{pmatrix}\boldsymbol{E}\\ \boldsymbol{H}\end{pmatrix} = \widehat{L}\left(\nabla, \frac{1}{c}\frac{\partial}{\partial t}\right)\begin{pmatrix}\boldsymbol{E}\\ \boldsymbol{H}\end{pmatrix} = \frac{4\pi}{c}\begin{pmatrix}\boldsymbol{j}_e\\ \boldsymbol{j}_m\end{pmatrix}, \tag{3}$$

with the operator $\widehat{Q} = \begin{bmatrix}\hat{0} & \widehat{\nabla}\times\widehat{I}\\ -\widehat{\nabla}\times\widehat{I} & \hat{0}\end{bmatrix}$.

One common way to express the fields generated by sources is to use the Green's function $\widehat{G}$

$$\begin{pmatrix}\boldsymbol{E}\\ \boldsymbol{H}\end{pmatrix} = 4\pi i k_0\,\widehat{G}\begin{pmatrix}\boldsymbol{j}_e\\ \boldsymbol{j}_m\end{pmatrix}, \quad \widehat{G} = i\omega\widehat{L}^{-1} \tag{4}$$

Alternatively, the solutions to Eq. (3) can be obtained using the ॐ-potential approach [2]

$$\begin{pmatrix}\boldsymbol{j}_e\\ \boldsymbol{j}_m\end{pmatrix} = D(\partial_x,\partial_y,\partial_z)\text{ॐ}(\boldsymbol{r}), \quad \begin{pmatrix}\boldsymbol{E}\\ \boldsymbol{H}\end{pmatrix} = \frac{4\pi}{c}\widehat{U}(\partial_x,\partial_y,\partial_z)\,\text{ॐ}(\boldsymbol{r}) \tag{5}$$

where the ॐ-potential is represented by $\text{ॐ}(\boldsymbol{r}) = \begin{pmatrix}\text{ॐ}_e\\ \text{ॐ}_m\end{pmatrix}$.

The operators are defined via their Fourier representation:

$$D(\partial_x,\partial_y,\partial_z) \to D(\boldsymbol{k}) = \det\widehat{L}(i\boldsymbol{k},-ik_0) \tag{6}$$

$$\widehat{U}(\partial_x,\partial_y,\partial_z) \to \widehat{U}(\boldsymbol{k}) = \operatorname{adj}\widehat{L}(i\boldsymbol{k},-ik_0) \tag{7}$$

Finally, in the momentum space, the Green's function is related to the ॐ-potential approach as

$$\hat{G}(i\mathbf{k}, -ik_0) = i\omega \frac{\text{adj } \hat{L}(i\mathbf{k}, -ik_0)}{\det \hat{L}(i\mathbf{k}, -ik_0)} \tag{8}$$

### 3. Beyond Light Shells

Fresnel wave surfaces represent the set of points in momentum space corresponding to plane waves that can propagate in a material at a given frequency. These surfaces are also known as *isofrequency surfaces*. In this work, we adopt the term *light shells* to emphasize the analogy with *mass shells* for massive particles in quantum field theory. This terminology allows us to classify far-field states located on the Fresnel wave surfaces as *on-shell waves*, while states with real k-vectors not belonging to the Fresnel surface are regarded as *off-shell waves*.

Mathematically, Fresnel wave surfaces are defined by the dispersion condition

$$D(\mathbf{k}) = \det \hat{L}(i\mathbf{k}, -ik_0) = 0$$

As an example, Fig. 2(a) shows a light shell corresponding to the material parameters matrix $\hat{M}$ depicted in Fig. 2(b).

Since $D = 0$ for on-shell far-fields, the operator $\hat{U}$ becomes singular, and all of its columns are proportional to the field amplitude eigenvector $(\mathbf{E}, \mathbf{H})$ of the corresponding plane wave. This behavior is illustrated in Fig. 2(c), where the dependence of $D$ on $k_z$ is plotted for the same material as in Figs. 2(a)–(b). In Fig. 2(c), we plot the $E_x$ component of all six columns of the $\hat{U}$ matrix, which coincide when $k_z$ corresponds to the on-shell condition $D = 0$.

Light shells determine not only the phase of far-field waves but also their polarization. The wave vectors $\mathbf{k}$ lying on the light shells are aligned with the Minkowski momentum, whose density is

$$g_M \propto \text{Re}\{\mathbf{D}^* \times \mathbf{B}\}.$$

This alignment arises because far-field solutions exclude external charges, enforcing the transversality conditions $k_0 \mathbf{D} = -\mathbf{k} \times \mathbf{H}$ and $k_0 \mathbf{B} = \mathbf{k} \times \mathbf{E}$. Normals to the light shells, on the other hand, are directed along the Abraham momentum

$$g_M \propto \text{Re}\{\mathbf{E}^* \times \mathbf{H}\},$$

whose misalignment with respect to the k-vectors reflects the generation of bound charges in isotropy-broken media [21, 22]. An example of this Minkowski–Abraham momentum locking is shown in Fig. 2(a) for a plane wave with k parallel to the z-axis.

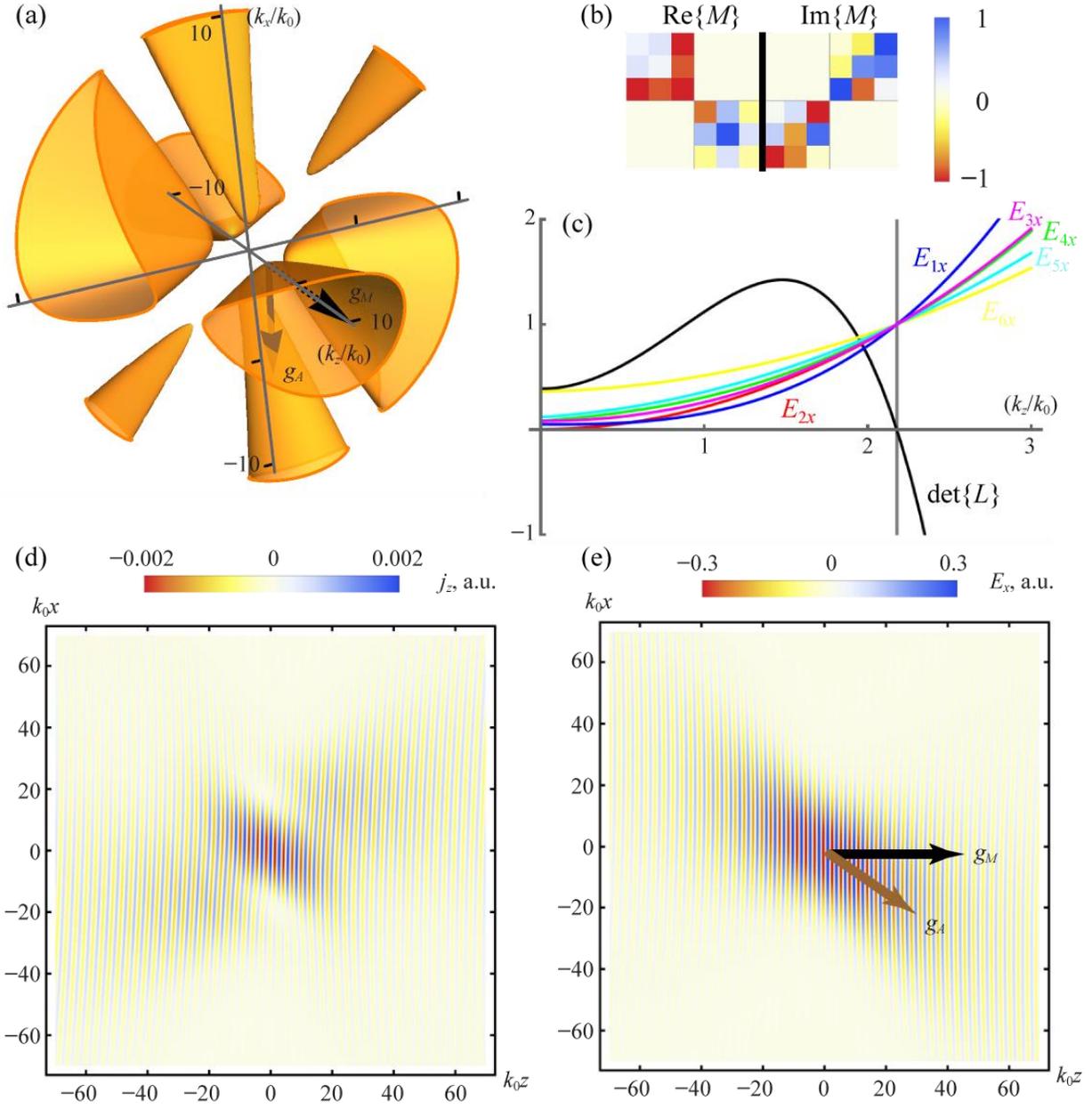

Fig. 2. (a) Fresnel wave surface (light shell) corresponding to the material parameters matrix $\hat{M}$ shown in panel (b). The directions of the Minkowski momentum $\boldsymbol{g}_M$ and Abraham momentum $\boldsymbol{g}_A$ are indicated for a representative on-shell plane wave. (b) Real and imaginary parts of the material parameters matrix $\hat{M}$ used to generate the surface in panel (a). (c) Dependence of the dispersion determinant $D$ on $k_z$ for the same material. Also shown are the $E_x$ components of all six columns of the operator $\hat{U}$; these components become identical at the on-shell condition $D = 0$. (d) External source distribution $j_x$ exciting the medium in paraxial approximation. (e) Resulting $E_x$ field of a Gaussian beam composed of on-shell states in the same approximation. The beam propagates along the Abraham momentum axis $\boldsymbol{g}_A$, illustrating Minkowski–Abraham momentum locking.

The Minkowski–Abraham momentum locking influences both fundamental and structured beams in isotropy-broken media, giving them a characteristic biaxial nature. This effect is illustrated in Figs. 2(d)-(e). As a first example, we consider a Gaussian beam represented in terms of the Om-potential:

$$\breve{\Omega} = \begin{pmatrix} \breve{\Omega}_e(\boldsymbol{r}) \\ \breve{\Omega}_m(\boldsymbol{r}) \end{pmatrix} e^{i(\boldsymbol{k}\boldsymbol{r} - k_0 ct)}$$

where $\breve{\Omega}_e(\boldsymbol{r})$ and $\breve{\Omega}_m(\boldsymbol{r})$ denote the beam envelopes. The spatial harmonics of such a beam lie on the light shell in the paraxial approximation. Due to Minkowski–Abraham momentum locking, the phase of the beam propagates along the Minkowski axis, whereas the beam ray itself is directed along the Abraham axis, consistent with Ref. [21].

In the paraxial approximation, the spatial harmonics of the beam correspond to the parabola tangent to the Fresnel wave surface, obtained from the Taylor expansion of the dispersion relation $D(\boldsymbol{k}) = 0$

$$D(\boldsymbol{k} + \delta\boldsymbol{k}) = \det \hat{L}(\boldsymbol{k} + \delta\boldsymbol{k}) \approx C\, \delta\boldsymbol{k} \cdot \boldsymbol{g}_{Abr} + \frac{1}{2!}\, \delta\boldsymbol{k} \cdot \hat{H} \cdot \delta\boldsymbol{k} = 0$$

where constant $C = \frac{1}{g_{Abr}} \frac{\partial D}{\partial k}$, $\boldsymbol{g}_{Abr}$ is the Abraham momentum density, and $\hat{H}$ is the Hessian matrix of the Fresnel wave surface at the central $\boldsymbol{k}$-vector.

In Fig. 2(d) we show the distribution of external sources corresponding to the $\breve{\Omega}$-potential with the same central wave vector $\boldsymbol{k}$ as discussed in Fig. 2(a). These sources are small in magnitude and arise from the deviation of the paraxial approximation from the exact solution of Maxwell's equation given by Eq. (5). The associated $E_x$ field distribution of this biaxial Gaussian beam is presented in Fig. 2(e). The field differs from the exact homogeneous Maxwell solution only because of the paraxial approximation and is therefore representative of the far-field regime.

To investigate near-fields, we turn to the off-shell spatial harmonics. Departing from the paraxial approximation, we construct the beam from spatial harmonics lying in the plane perpendicular to the Abraham momentum at wave vector $\boldsymbol{k}$, i.e. satisfying

$$\delta\boldsymbol{k} \cdot \frac{\partial D(\boldsymbol{k})}{\partial \boldsymbol{k}} = 0$$

In this regime the external sources can no longer be neglected, and Eq. (9) applies:

$$\begin{pmatrix} \boldsymbol{j}_e(\boldsymbol{r}) \\ \boldsymbol{j}_m(\boldsymbol{r}) \end{pmatrix} = D(\partial_x, \partial_y, \partial_z) \breve{\Omega}(\boldsymbol{r}) \approx \frac{1}{2!}(-i\nabla) \cdot \hat{H} \cdot (-i\nabla) \begin{pmatrix} \breve{\Omega}_e(\boldsymbol{r}) \\ \breve{\Omega}_m(\boldsymbol{r}) \end{pmatrix} + \cdots \quad (9)$$

The external sources defined by Eq. (9) are illustrated in Fig. 3(a) (in the x–z plane) and Fig. 3(c) (cross-section in the x–y plane at $z = 0$). One can see that the wavefronts in this case are flat.

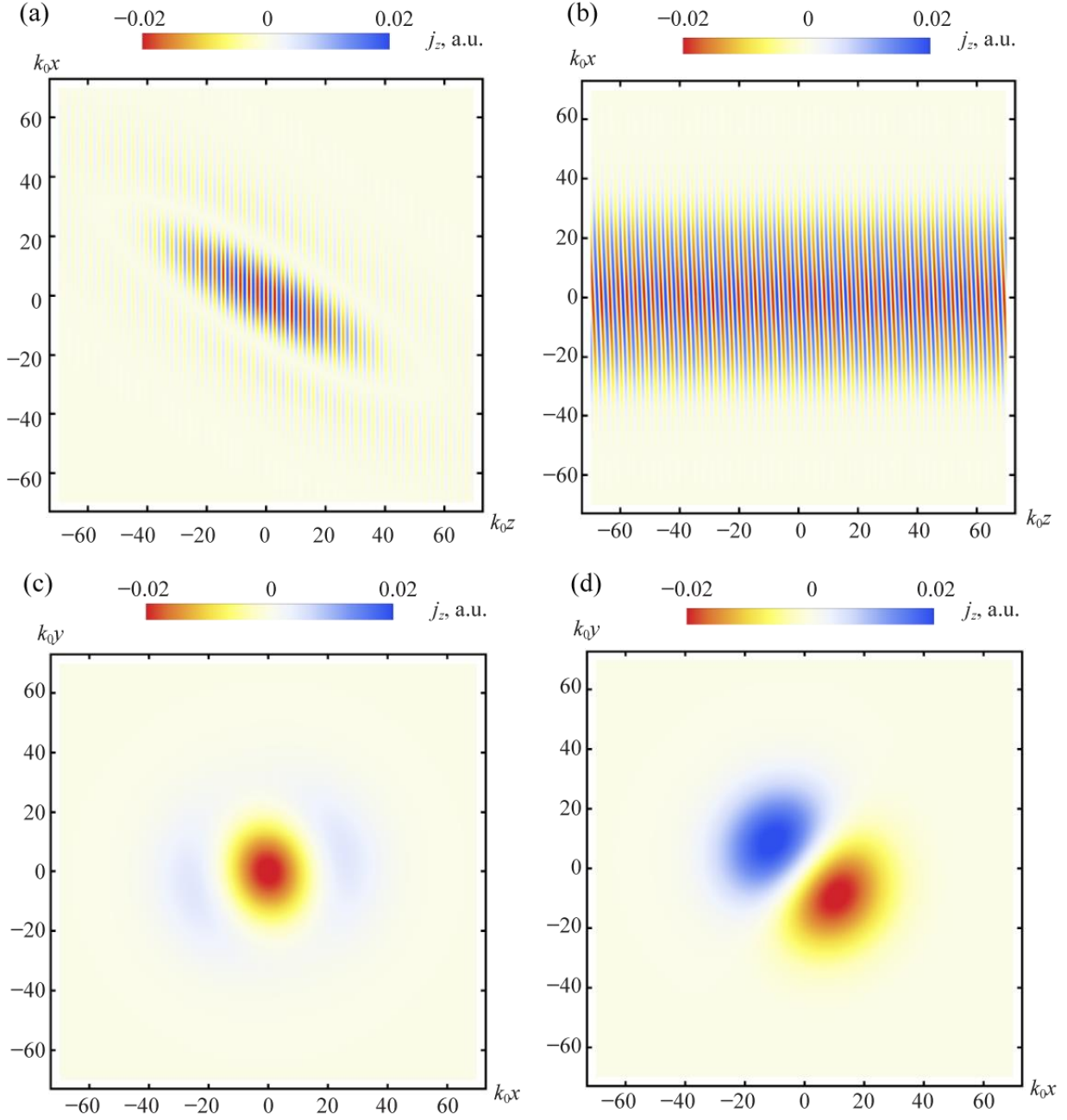

Fig. 3. External sources defined by Eq. (9) (a) in x-z plane and (c) in x-y plane at $z = 0$; External sources given by Eq. (10) (b) in x-z plane and in x-y plane (c) at $z = 0$; (d) at $z = \pi/(2k_z)$.

In the more general case of a Gaussian beam centered at a wave vector $\boldsymbol{k}$ on the light shell, the sources are

$$\begin{pmatrix} \boldsymbol{j}_e(\boldsymbol{r}) \\ \boldsymbol{j}_m(\boldsymbol{r}) \end{pmatrix} \approx C \; \boldsymbol{g}_{Abr} \cdot (-i\nabla) \begin{pmatrix} \breve{\mathcal{P}}_e(\boldsymbol{r}) \\ \breve{\mathcal{P}}_m(\boldsymbol{r}) \end{pmatrix} + \frac{1}{2!}(-i\nabla)\cdot \widehat{H} \cdot (-i\nabla) \begin{pmatrix} \breve{\mathcal{P}}_e(\boldsymbol{r}) \\ \breve{\mathcal{P}}_m(\boldsymbol{r}) \end{pmatrix} + \cdots \quad (10)$$

-7-

This expression clarifies the role of the Abraham momentum. In the far-field regime, it determines the direction of beam propagation. By contrast, for off-shell spatial components whose k-vectors deviate from the Fresnel wave surface, the Abraham momentum is no longer a property of the field itself but instead characterizes the external sources. In this picture, sources are expressed as directional derivatives of the ॐ-potential envelope along the Abraham momentum direction.

For a Gaussian beam composed of spatial harmonics lying in the plane tangent to the Minkowski momentum condition $\delta \boldsymbol{k} \cdot \boldsymbol{k} = 0$, the external sources defined by Eq. (10) are shown in Fig. 3(b) in the x–z plane and Fig. 3(c) in the x–y plane at $z = 0$, and in Fig. 3(d) at $z = \pi/(2k_z)$. Note that the cross-section at $z = 0$ is the same for the sources described by Eqs. (9) and (10). The additional first term in Eq. (10) modifies the cross-section at $z = \pi/(2k_z)$, producing the distribution shown in Fig. 3(d). It should be emphasized that the first and second terms in Eq. (10) are out of phase.

The resulting field distributions produced by these sources are presented in Fig. 4. Since all of the spatial harmonics in these pulses remain close to the Fresnel wave surface, the $\widehat{U}$ matrices are still ill-conditioned, and the polarization is only weakly perturbed from its far-field direction at the Fresnel surface. The perturbation depends only slightly on the orientation of the ॐ-potential.

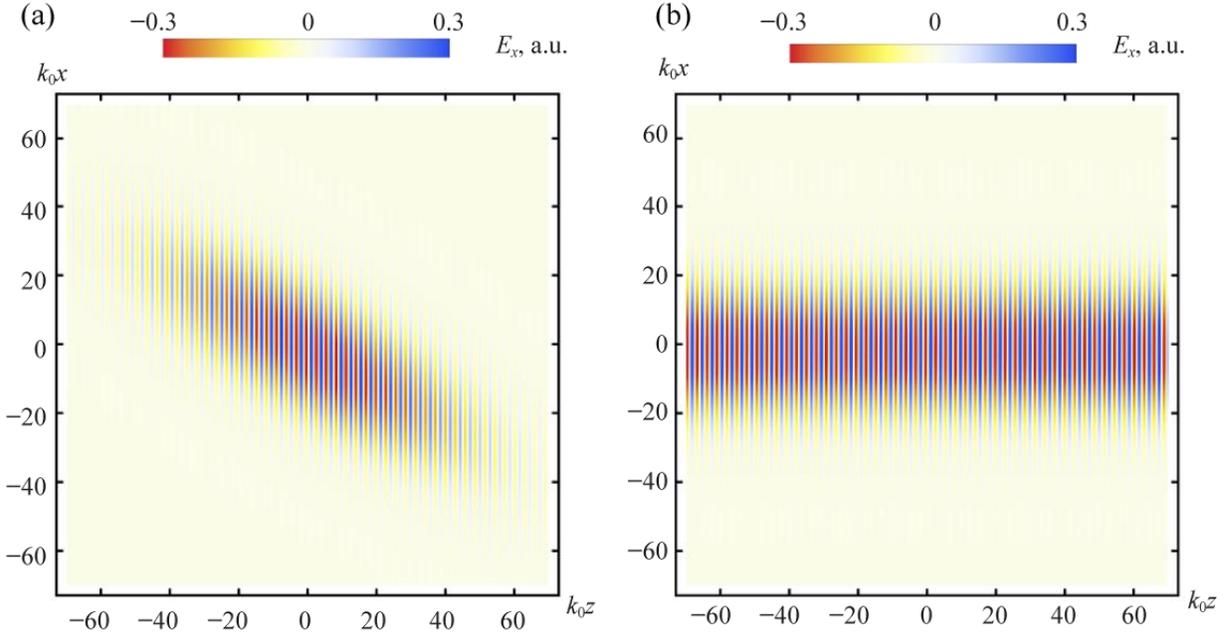

Fig. 4. Distribution of the $E_x$ component corresponding to the external sources shown in Fig. 3(a)–(b), calculated using Eq. (5).

If the spatial harmonics of a beam with central wave vector $\boldsymbol{k}$ lie far from the Fresnel wave surface in momentum space, then $D(\boldsymbol{k})$ is nonzero and acts as a scalar proportionality factor between the external sources and the ॐ-potential:

$$\begin{pmatrix}j_e(r)\\j_m(r)\end{pmatrix} \approx D(k)\,\breve{\mathcal{P}}(r) = D(k)\begin{pmatrix}\breve{\mathcal{P}}_e(r)\\\breve{\mathcal{P}}_m(r)\end{pmatrix}e^{i(kr-k_0ct)}$$

This relation shows that the orientation and type of external sources are dictated by the polarization of the $\breve{\mathcal{P}}$-potential. In this case, both the $\hat{U}$ and $\hat{L}$ operators are well-conditioned, and the electromagnetic fields depend explicitly on the external sources:

$$\begin{pmatrix}E\\H\end{pmatrix} = \frac{4\pi}{c}\frac{ik_0}{c}\hat{U}(k)\,\breve{\mathcal{P}}(r) = \frac{4\pi}{c}\hat{L}^{-1}(k)\begin{pmatrix}j_e(r)\\j_m(r)\end{pmatrix}$$

### 4. Beyond Reciprocal Media: Lorentz Broadening of Light Shells and BBL

In reciprocal materials, Fresnel wave surfaces are well defined by the dispersion condition $D = \det\hat{L} = 0$. By contrast, the Fresnel wave surfaces of non-reciprocal media are the subject of active research in modern optics [25]. The difficulty lies in the fact that, for non-reciprocal materials, $D = \det\hat{L}$ is a complex quantity [26]. As a result, the conditions $\mathrm{Re}\{\det\hat{L}\} = 0$ and $\mathrm{Im}\{\det\hat{L}\} = 0$ define curves rather than a single closed surface in the space of real wave vectors $k$ [27].

In many treatments, one therefore considers complex wave vectors $k = (k' + ik'')\hat{k}$ that satisfy $D = \det\hat{L} = 0$. In this case, the Fresnel wave surface is represented in the momentum space of the real parts $k = k'\hat{k}$, while the imaginary parts are not investigated [25].

The imaginary components of $k$ play a crucial role in energy balance during plane-wave propagation in non-reciprocal media with loss or gain. This results in exponential attenuation or amplification of electromagnetic fields - a process captured by the Beer–Bouguer–Lambert (BBL) law. While the BBL law has long been recognized in many branches of science, its implications for isotropy-broken media have not been systematically studied.

In what follows, we develop a unified framework for Fresnel wave surfaces in non-reciprocal media. The approach models Lorentzian broadening of light shells, explicitly incorporates BBL-type exponential modulation, and keeps the description in real k-space.

From Maxwell's equations, we obtain the following relationship

$$k_0(E^*\cdot D + H^*\cdot B) + \frac{4\pi i}{c}(E^*\cdot j_e + H^*\cdot j_m) = 2\,k\cdot\mathrm{Re}\{E^*\times H\} \qquad (11)$$

Introducing the bound currents,

$$j_{be} = \frac{\partial P}{\partial t} = -i\omega P,\quad j_{bm} = \frac{\partial M}{\partial t} = -i\omega M$$

the material absorption (work done on bound charges/currents) can be written as

$$Q = k_0\mathrm{Im}\{E^*\cdot D + H^*\cdot B\} = \frac{4\pi}{c}\mathrm{Re}\{E^*\cdot j_{be} + H^*\cdot j_{bm}\}.$$

The power exchanged with external sources is

$$P = \frac{4\pi}{c} \text{Re}\{\boldsymbol{E}^* \cdot \boldsymbol{j}_e + \boldsymbol{H}^* \cdot \boldsymbol{j}_m\}.$$

Taking the imaginary part of Eq. (11) yields

$$P = Q + 2 \text{ Im } \boldsymbol{k} \cdot \text{Re}\{\boldsymbol{E}^* \times \boldsymbol{H}\},$$

which shows that the energy delivered by sources during one optical period is partitioned into: (i) local work $Q$ of electromagnetic fields on bound currents in the material (true absorption/gain), and (ii) a term proportional to Im $\boldsymbol{k}$, which produces exponential attenuation or amplification of the fields - i.e., the Beer–Bouguer–Lambert effect.

In reciprocal, lossless media the field–material exchange is purely reactive, so $Q = 0$. If, additionally, there is no decay or gain (Im $\boldsymbol{k} = 0$), the exchange with external sources is also purely reactive, $P = 0$. This conclusion does not apply on the Fresnel wave surface where $D = \det \hat{L} = 0$, and therefore $\hat{L}^{-1}$ diverges [see Eq. (8)]; any nonzero external current then drives fields that diverge.

This gives a clue for generalizing the Fresnel wave surface concept to non-reciprocal media. Consider an external electric source polarized along $\hat{\boldsymbol{x}} : \boldsymbol{j}_e = j_e \hat{\boldsymbol{x}}$. The energy exchanged between the field and the source is

$$P = \frac{4\pi}{c} \text{Re}\{\boldsymbol{E}^* \cdot \boldsymbol{j}_e\} = \left(\frac{4\pi}{c}\right)^2 \omega \text{ Im}\{\boldsymbol{j}_e \cdot \hat{G} \cdot \boldsymbol{j}_e^*\} = \left(\frac{4\pi}{c}\right)^2 \omega |j_e|^2 \text{Im}\{\hat{G}_{xx}\} = 8\pi^3 |j_e|^2 \rho$$

Here the photonic density of states (PDOS) for the x-polarization is

$$\rho = \frac{2\omega}{\pi c^2} \text{Im}\{\hat{G}_{xx}\} = \frac{2\omega^2}{\pi c^2} \text{Re}\left\{\frac{(\text{adj } \hat{L})_{xx}}{\det \hat{L}}\right\},$$

using $\hat{G} = i\omega \hat{L}^{-1} = i\omega(\text{adj } \hat{L})/(\det \hat{L})$.

From Eq. (5) we have

$$(\text{adj } \hat{L})_{xx} = \frac{c E_x}{4\pi \check{\wp}_{ex}} = i\alpha.$$

The overall field phase can be chosen so that $\alpha$ real.

Now consider the PDOS near the Fresnel wave surface, where $k \approx k_0 n$. Linearizing,

$$\det \hat{L} \approx A(k - k_0 n)$$

With the renormalization $\alpha = c^2 A \sigma / (2\omega^2)$, we get

$$\rho = \frac{\sigma}{\pi} \text{Im}\left\{\frac{1}{k - k_0 n}\right\}.$$

where the surface density of states (SDOS) is

$$\sigma = \frac{2\omega^2}{ic^2} \frac{(k - k_0 n)}{\det \hat{L}} \left(\text{adj } \hat{L}\right)_{xx} = \frac{2\omega^2}{ic^2} (k - k_0 n) \left(\hat{L}^{-1}\right)_{xx} \quad \text{at } D = 0$$

Intuitively, $A = \partial(\det \hat{L})/\partial k$ is the slope of $\det \hat{L}$ along the surface normal, while $-i\left(\text{adj } \hat{L}\right)_{xx}$ sets the Green-function residue for the chosen polarization; their ratio thus defines the surface density of states $\sigma$ on the Fresnel wave surface.

To include non-reciprocity, perturb the constitutive matrix

$$\hat{M}' = \hat{M} + \delta\lambda \cdot \hat{\Lambda}, \quad \delta\lambda \ll 1$$

so that in Fourier space $\hat{L}' = \hat{Q}(i\mathbf{k}) - ik_0 \hat{M}' = \hat{L} + ik_0 \delta\lambda \cdot \hat{\Lambda}$.

Using Jacobi's formula,

$$\det \hat{L}' = \det \hat{L} + ik_0 \delta\lambda \cdot \text{tr}\left(\text{adj}(\hat{L}) \cdot \hat{\Lambda}\right)$$

Near the Fresnel surface we obtain

$$\frac{A}{\det \hat{L}'} \approx \frac{1}{k_r - k_0(n + i\kappa)}$$

where $(n + i\kappa)$ is the complex index of refraction corresponding to a solution of $\det \hat{L}' = 0$. Therefore, in media with gain or loss the polarization-specific PDOS becomes

$$\rho = \frac{\sigma}{\pi} \text{Im}\left\{\frac{1}{k_r - k_0(n+i\kappa)}\right\} = \frac{\sigma}{\pi} \frac{k_0 \kappa}{(k_r - k_0 n)^2 + (k_0 \kappa)^2} = \sigma\, L(k_r - k_0 n, 2k_0 \kappa) \tag{12}$$

Here $L$ is the Lorentzian function. Note that in reciprocal materials $\kappa \to 0$ and $L \to \delta(k_r - k_0 n)$ giving a delta-functional density of states peaking on the Fresnel wave surface. As seen from Eq. (12), this relates Fresnel wave surfaces directly to the PDOS in momentum space and shows that non-reciprocity produces a Lorentzian broadening of the surface in PDOS. The width of this broadened PDOS is proportional to the imaginary part of the direction-dependent refractive index $n + i\kappa$, which links the broadening to direction-dependent BBL extinction or amplification of electromagnetic waves in non-reciprocal media.

To clarify the meaning of our results, consider the case of amplification. The BBL modulation then has the form $e^{ik_0(n-i|\kappa|)z}$, which corresponds to a negative $\kappa$. By Eq. (12) this produces a negative PDOS $\rho$. Because the source–field power exchange is $P = 8\pi^3 |j_e|^2 \rho$, a negative $\rho$ implies $P < 0$: the electromagnetic field returns active power to the source.

In real k-space we set Im $\mathbf{k} = 0$, so we have $Q = P$. Thus $Q < 0$ identifies gain: the material does work on the field, and the field delivers power back to the source. By contrast, for loss ($\kappa > 0$) we have $\rho > 0$ and $P = Q > 0$, which is the usual situation where power flows from the source to the field and is then dissipated in the material.

In the reciprocal, lossless case $\kappa = 0$, the BBL effect vanishes, and the exchange is purely reactive, so $P = Q = 0$ over one optical period. The PDOS collapses to a delta peak on the Fresnel surface, reflecting an undamped on-shell mode. Exactly on the shell the Green's function is singular, so any finite source would diverge in the idealized model; in practice, small damping always regularize this to a narrow, symmetric Lorentzian.

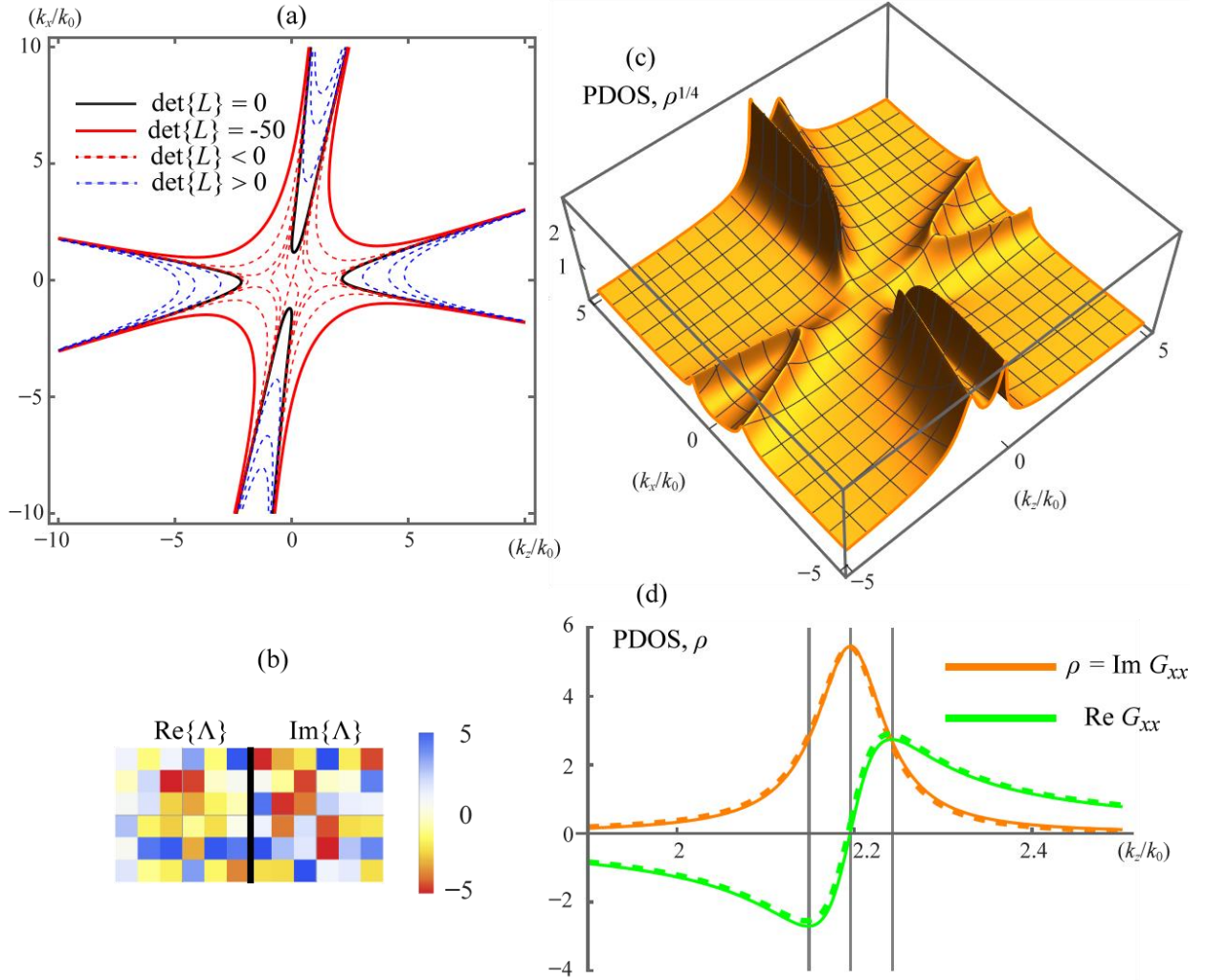

Fig. 5. (a) $k_x$-$k_z$ cross-section of the light shell from Fig. 2(a). Shown are the on-shell contour $\det \hat{L} = 0$ (black) together with representative level sets for $\det \hat{L} > 0$ and $\det \hat{L} < 0$ (line styles indicate the sign). (b) Real and imaginary parts of the constitutive matrix $\hat{\Lambda}$ used for panels (c)–(d) (color-coded). (c) PDOS map $\rho^{1/4}$ in $k_x$-$k_z$ plane for the non-reciprocal medium defined by $\hat{M}$ in Fig. 2(b) and $\hat{\Lambda}$ in (b) with $\delta\lambda = 10^{-3}$ computed from $\rho \propto \text{Im}\{G_{xx}\}$. (d) Lineout of the PDOS $\rho \propto \text{Im}\{G_{xx}\}$ (orange) and $\text{Re}\{G_{xx}\}$ (green) versus $(k_z/k_0)$ (solid) along the $k_z$-axis in (c) and the corresponding Lorentzian fit near the Fresnel wave surface (dashed); The vertical bars mark the peak position $k_z = k_0 n$ (center) and the full width at half maximum $k_z = k_0(n \pm \kappa)$ (side bars), evidencing the Lorentzian broadening of the Fresnel wave surface predicted by Eq. (12).

## 5. Conclusion

We have extended the classical framework of Fresnel wave surfaces to encompass near-field regimes, non-reciprocal responses, and media with loss or gain. By employing the Om-potential approach, we established a unified description that treats far-field states as *on-shell* modes and near-field contributions as *off-shell* modes, directly paralleling concepts from quantum field theory. This interpretation clarifies the role of Abraham and Minkowski momenta, which in far-field propagation are locked to light shells but, in the near-field regime, emerge as descriptors of source structure.

A key result of this work is the demonstration that photonic density of states (PDOS) in non-reciprocal and non-Hermitian media exhibits Lorentzian broadening around Fresnel wave surfaces. This broadening provides a direct connection between the Beer–Bouguer–Lambert law of exponential attenuation or amplification and the momentum-space topology of electromagnetic fields. In this way, we bridge a theoretical gap by linking the physics of sources, near-fields, and structured radiation to established tools of dispersion theory.

Beyond its conceptual significance, the proposed framework offers practical value for the design of emitters, metamaterials, and structured-light systems. It suggests new approaches for tailoring PDOS and source-field interactions in anisotropic, hyperbolic, and non-Hermitian photonic platforms. Such advances are directly relevant to emerging applications including 6G telecommunications, multiplexing via structured light, and topological photonics.

Future work may expand these ideas toward explicit experimental verification, numerical modeling of PDOS broadening in specific material systems, and integration with non-Hermitian photonic devices. By reconciling on-shell and off-shell modes, near- and far-fields, and reciprocal and non-reciprocal responses, this work opens pathways to both deeper fundamental understanding and new technological frontiers in photonics.

**Acknowledgements.** The authors acknowledge the support from the Center for Advanced Materials Science at Georgia Southern University. Fig. 1 was drawn by Katie Durach.